\documentclass[10pt,article,superscriptaddress,amsmath,amssymb,aps,pra, longbibliography,floatfix]{revtex4-1}

\usepackage[T1]{fontenc}
\usepackage[utf8]{inputenc}
\usepackage{tgpagella}
\usepackage{gensymb}
\usepackage{setspace}
\usepackage{graphicx}
\usepackage{xcolor}
\usepackage{bm}
\usepackage{tabularx}
\usepackage{soul}
\soulregister\cite7
\usepackage{xcolor}
\sethlcolor{yellow}

\usepackage{comment}
\usepackage{enumitem}

\usepackage{textcomp,mathcomp}

\usepackage{color}
\usepackage{dcolumn}

\usepackage[english]{babel}
\usepackage[bottom=2.5cm,top=2.5cm, left=2cm, right=2cm]{geometry}
\usepackage{siunitx}
\usepackage{xr}
\makeatletter
\newcommand*{\addFileDependency}[1]{
  \typeout{(#1)}
  \@addtofilelist{#1}
  \IfFileExists{#1}{}{\typeout{No file #1.}}
}
\makeatother

\usepackage[unicode=true, bookmarks=true, bookmarksnumbered=false, bookmarksopen=false, breaklinks=false, pdfborder={0 0 1}, backref=false, colorlinks=false]{hyperref}

\begin{document}
\title{Manipulating ferroelectricity without electrical bias: A perspective}

\author{Bixin Yan}
\email[]{bixin.yan@mat.ethz.ch}
\affiliation{Department of Materials, ETH Zurich, CH-8093 Zurich, Switzerland}
\author{Valentine Gillioz}
\affiliation{Department of Materials, ETH Zurich, CH-8093 Zurich, Switzerland}
\author{Ipek Efe}
\affiliation{Department of Materials, ETH Zurich, CH-8093 Zurich, Switzerland}
\author{Morgan Trassin}
\email[]{morgan.trassin@mat.ethz.ch}
\affiliation{Department of Materials, ETH Zurich, CH-8093 Zurich, Switzerland}

\date {\today}

\begin{abstract}
Ferroelectric materials are established candidates for beyond complementary metal-oxide-semiconductor technology, owing to their non-volatile spontaneous electrical polarization. The recent boom in electric dipole texture engineering and manipulation in such materials has revealed exciting routes for controlling ferroelectric polarization, offering alternatives to the classical, sometimes challenging, application of electrical fields. In this short perspective, we shed light on electrode-free external stimuli enabling control over polar states in thin films. We bring awareness to the polarizing role of chemically-engineered surface contributions and provide insights into the combination of chemical substitution and mechanical pressure, complementing the polar state tuning capabilities readily enabled by flexoelectricity. Finally, we describe recent developments in the optical modulation of polarization. Thus, our perspective aims to stimulate the advancement of alternative means to act on polarization states and facilitate the development of ferroelectric-based applications.

\end{abstract}
\maketitle

Materials hosting a spontaneous electrical polarization, which can be reversed upon the application of an electric field, are referred to as ferroelectrics \cite{setter2006ferroelectric}. The low-energy cost accompanying voltage application and the compatibility of electric fields with down-scaling render such systems extremely interesting for the next generation of oxide electronics for beyond complementary metal-oxide-semiconductor technology and memory-in-logic applications \cite{manipatruni2018beyond,manipatruni2019scalable,muller2023ferroelectric}. The lattice deformation resulting from the intrinsic polar displacement in ferroelectrics makes all ferroelectrics piezoelectrics and further extends the industrial value of such materials with their dominant impact in transducer \cite{cross_ferroelectric_1996}, actuator, and sensor technologies \cite{muralt2000ferroelectric}. Numerous reviews in the literature have extensively covered the fundamentals and the application relevance of ferroelectrics \cite{damjanovic1998ferroelectric,setter2006ferroelectric, rodel2009perspective,scott2007applications,dawber2005physics,martin2016thin}
. Here, with this perspective, we will describe emerging routes for controlling polarization states in epitaxial oxide thin films, especially highlighting the development of electrical-bias-free manipulation of polar states. 
While our focus will primarily lie in showcasing the control on pre-existing polarization states in ferroelectric thin films, the enforcement of a polarization emergence in materials will be discussed.

In an effort to extend the control of polar textures at the nanoscale in oxide multilayers, harnessing the combination of structural, chemical, and electrostatic contributions at the films interfaces has recently enabled the creation of new electrical orders in matter \cite{strkalj2019design,susarla2021atomic}. The design of synthetic antiferroelectrics \cite{yin2024mimicking,catalan2025switching}, the emergence of ferrielectric-like order \cite{efe2025nanoscale}, and the stabilization of polar and nonpolar phase coexistence \cite{caretta2023non,muller2025reversible} are prominent examples of the most recent developments in the field. These interface-driven effects may exhibit unconventional responses to external stimuli. The demonstration of the optical creation of a supercrystal elegantly demonstrates the extended capabilities offered by synthetic polar systems \cite{stoica2019optical}. More generally, there is a need to identify and control electrical-bias-free means to act on the polarization to further extend technological developments enabled by the integration of ferroelectric materials into applications.

\begin{figure*}[hbt!]
    \includegraphics[width=.6\columnwidth]{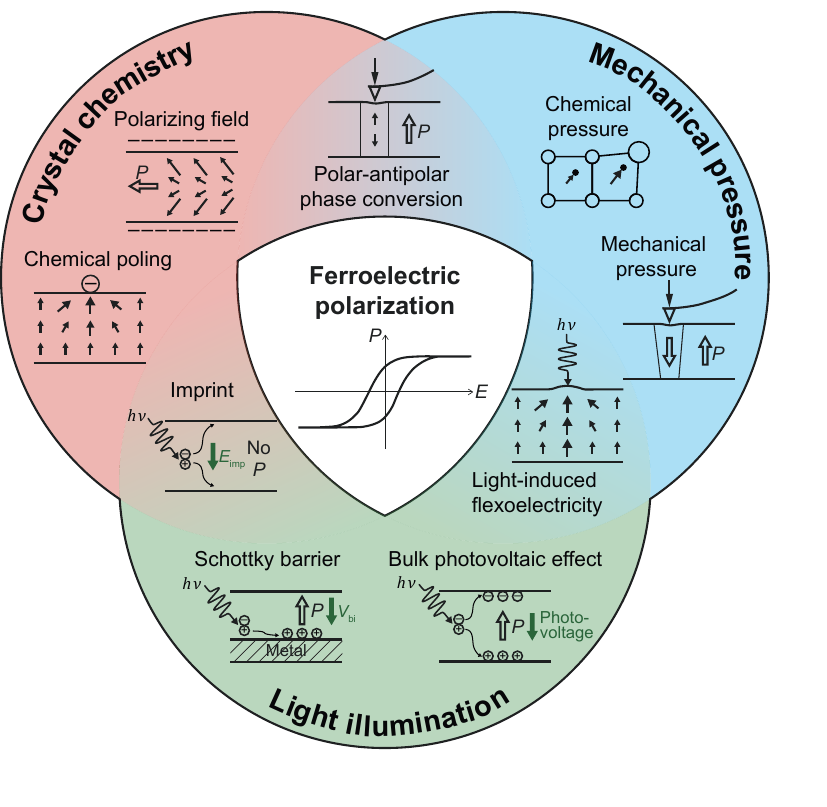}
    \caption{\textbf{Electrical-bias-free means for manipulating ferroelectric polarization}. Three routes employing crystal chemistry (top left), mechanical pressure (top right), and light illumination (bottom) are highlighted and schemed. The intersections show possible complementary mechanisms. Solid arrows in the films represent local electrical dipoles, and the hollow arrows represent the macroscopic ferroelectric polarization. $E_{\text{imp}}$ denotes the electric imprint field in the heterostructure and $V_{\text{bi}}$ denotes the built-in voltage near the interface. Both will be discussed in detail in section \ref{sec_l}.}
    \label{scheme}
    \end{figure*}

In this article, we present our perspective on the advances in the manipulation of polar states in thin films by electrical-bias-free means, as illustrated in \autoref{scheme}. 
We will first address the newest development regarding the use of polarizing surfaces to manipulate polar order in thin films in section \ref{sec_c}, as shown by the top left panel of \autoref{scheme}. This is in contrast with the most commonly employed depolarizing-tuning route. Hence, the use of charged, chemically engineered surfaces with defect ordering may provide new degrees of freedom in the design of synthetic polar states and ferroelectric poling.
In section \ref{sec_p}, we will cover the impact of local mechanical pressure application as a means of continuously controlling the net polarization state in thin films (\autoref{scheme} top right panel). We will highlight how pressure-driven polar phase conversion can add to the exciting field of flexoelectricity. 
Finally, we will discuss the light-matter interaction in ferroelectric thin films and the optical manipulation of ferroelectricity to advance the understanding of remote optical control of polarization in section \ref{sec_l}, see the bottom panel in \autoref{scheme}.

\section{Controlling electric dipole orientations using polarizing surfaces and crystal chemistry}
\label{sec_c}

    \begin{figure*}[hbt!]
    \includegraphics[width=\columnwidth]{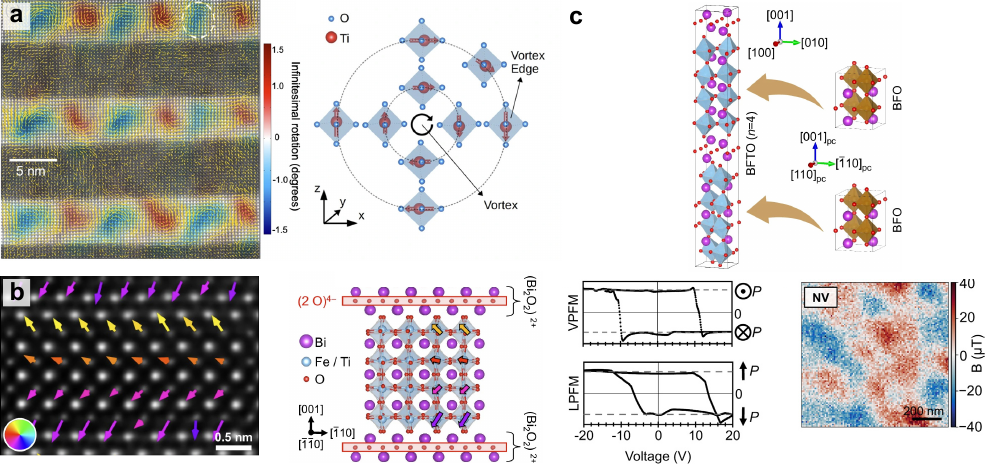}
    \caption{\textbf{Control electric dipole orientations using polarizing surface and crystal chemistry}. 
    (a) Polar vortices in PbTiO$_3$/SrTiO$_3$ (PTO/STO) superlattices. Left: $A$-site ($A$: Pb or Sr) displacement vectors (yellow arrow) and curl of displacement (red/blue color) overlaid on the high-angle annular darkfield scanning transmission electron microscopy (HAADF-STEM) image. The color bar indicates the magnitude of the curl of the displacement vector. Right: schematic representing the rotation of TiO$_6$ octahedra within one vortex domain. Reprinted with permission from Ref.\ \cite{susarla2021atomic}. Copyright 2021 by the Authors under Creative Commons Attribution 4.0  International License published by the Nature Publishing Group.
    (b) HAADF-STEM image with the measured electric dipole distribution overlaid (left) and schematic (right) of the Aurivillius Bi$_5$FeTi$_3$O$_{15}$ (BFTO) unit cell. Each half unit cell includes four perovskite layers and a Bi$_2$O$_2$ sheet. The negatively charged oxygen atomic layer in the Bi$_2$O$_2$ is highlighted in red. The arrows represent the electric dipoles pointing toward the nearest Bi$_2$O$_2$ layer. The arrows in the left panel are colored according to the given 360° color wheel and show the polarization vector at each B-site cation position.
    (c) Schematic of the insertion of BiFeO$_3$ (BFO) into the Aurivillius layered framework (top), local piezoresponse switching spectroscopy and scanning nitrogen vacancy (NV) magnetometry image of the composite film (bottom). The phase of the vertical piezoresponse force microscopy (VPFM) and lateral piezoresponse force microscopy (LPFM) signal recorded during the poling of the film reveals the existence of a net polarization along both the in-plane and out-of-plane directions. Scanning NV magnetometry image revealing antiferromagnetic domains. (b) and (c) are reprinted with permission from Ref.\ \cite{efe2025nanoscale}. Copyright 2025 by the Authors under Creative Commons Attribution 4.0  International License published by the Nature Publishing Group.
    }
    \label{chemistry}
    \end{figure*}

\textbf{Tuning the depolarizing field contribution in ferroelectric thin films.} Engineering of polarization states in ferroelectric thin films and superlattices has, so far, mostly been enabled by the tuning of the depolarizing field contributions using various buffer or spacer layer integrations \cite{lichtensteiger2011ferroelectricity,lichtensteiger2014tuning,yadav2016observation,strkalj2019design,strkalj2019depolarizing}. In ferroelectric materials, bound charge accumulation at the surface leads to the emergence of the so-called depolarizing field, which is oppositely oriented to the polarization \cite{junquera2003critical}. Acting on the screening of the ferroelectric surface-bound charge, therefore, allows balancing this depolarizing contribution and controlling the final polarization state \cite{strkalj2019depolarizing, trassin2022bringing}. Due to the polarization suppression mechanism in thin films, when ferroelectrics are interfaced with dielectric, non-charge-screening layers, oppositely oriented domain formation \cite{strkalj2019depolarizing}, polarization rotation \cite{yadav2016observation,susarla2021atomic}, and antipolar phases \cite{mundy2022liberating,caretta2023non,catalan2025switching} can be stabilized at remanence, as shown in \autoref{chemistry} (a). This approach has been instrumental in the advancement of complex electric dipole texture design. Polar vortices and skyrmions in PbTiO$_3$/SrTiO$_3$ (PTO/STO) ferroelectric/dielectric superlattices are prominent examples of the technological relevance of the degree of freedom offered by the manipulation of the depolarizing field contribution in oxide thin films \cite{yadav2016observation,susarla2021atomic,strkalj2022optical,das2019observation}.
Such polar textures represent stable, nanoscale topological structures, which open new opportunities for low-energy-consumption and high-density data storage \cite{junquera2023topological,chen2025emergence,geng2025dipolar,feng2025interfacial}.
More recently, the capacity to reversibly cancel and revive the remanent net polarization has been employed to create the first synthetic antiferroelectric system in the canonical PTO/STO superlattice system \cite{yin2024mimicking}. There, substrate selection enables a tensile epitaxial strain that enforces an in-plane configuration of the ferroelectric order at remanence \cite{yin2024mimicking,strkalj2021stabilization}. An electric field applied in the out-of-plane direction then triggers a polarization rotation along the field, resulting in the transition from a macroscopically nonpolar state in the vertical direction to a net polar configuration. Most strikingly, the depolarizing field contribution ensures the return to the initial in-plane polarized state once the field is removed, with the absence of an out-of-plane oriented remanent polarization. This all results in a double hysteresis loop in the polarization as a function of the electric field, characteristic of field-induced antipolar to polar phase conversion in antiferroelectric materials \cite{catalan2025switching}.

Nevertheless, despite its role in the engineering of electric dipole textures, the depolarizing field remains one of the major limitations in the integration of ferroelectric thin films in many technologically relevant applications \cite{junquera2003critical}. For ferroelectric memories \cite{scott1989ferroelectric}, ferroelectric tunnel junctions \cite{garcia2014ferroelectric}, or ferroelectric-field-effect transistors \cite{khan2020future,salahuddin2018era}, a robust remanent polarization is desired. The miniaturization of the device architecture leads to an increased depolarizing field contribution and, thus, a suppression of the polarization. 
Although significant progress has been made in combating thin-film depolarization \cite{nordlander2024combined, gradauskaite2023defeating,jiang2022enabling}, alternative approaches in electric dipole engineering may further extend the application relevance of ferroelectric multilayers with a non-zero net polarization at remanence.

\textbf{Polarizing layers as a new strategy for polarization engineering in ferroelectrics.} In contrast to the depolarizing field approach, the concept of using polarizing, charged interfaces in ferroelectric thin film architectures to drive the electric dipole orientation is gathering increased interest \cite{efe2024engineering}. The seminal work of Yu et al.\ highlighted the ability to deterministically control the polarization orientation in out-of-plane polarized thin films using the charged atomic plane termination of buffer layers \cite{yu2012interface}. There, the use of well-defined charged atomic surface terminations of oxide buffer layers enabled the creation of electrostatic potential steps across the interface, which set the preferred polarity of the bound charge accumulation and, hence, the final out-of-plane polarization direction in the ferroelectric layer grown on top \cite{de2017nanoscale,strkalj2020situ,gradauskaite2025magnetoelectric}. The polarity of the charged interface depends on the atomic plane sequences and the chemistry of the materials, which is also referred to as layer polarization in perovskite oxides \cite{efe2021happiness}.

Beyond this directional control, surface chemistry can be employed to locally induce a polarizing field and enhance the polarization in ferroelectric systems as an alternative approach to the dominant depolarizing-field tuning method in the design of unconventional electric dipole textures. Off-stoichiometric charged surface layers may also spontaneously form during the epitaxial growth of ferroelectric films and locally enhance or drive the orientation of the electric dipoles in the entire volume of the films \cite{strkalj2020situ,xie2017giant,gradauskaite2026control}. Most interestingly, in the family of layered ferroelectric compounds, which were first considered for ferroelectric memory applications, the unit cell consists of periodically repeating charged sheets interleaved with perovskite building blocks \cite{gradauskaite2025revival}. Remarkably, the succession of fluorite-like Bi$_2$O$_2$ charged sheets in the unit cell of layered Aurivillius compounds \cite{WOS:A1950XY97500010,gradauskaite2020robust,faraz2018exploring} leads to a splitting of the electric dipole orientation in the upward and downward directions, i.e., into a net antipolar ordering in the direction orthogonal to the layered architecture, as shown in \autoref{chemistry} (b) \cite{campanini2019buried, gradauskaite2020robust,moore2022charged,efe2025nanoscale}.

The structural compatibility of functional perovskite with most layered architectures calls for the investigation of composite design beyond the realm of the pure Aurivillius layered framework \cite{gradauskaite2023defeating,efe2025nanoscale}. The insertion of various ferroelectric perovskite constituents within the  Aurivillius layered framework, such as multiferroic BiFeO$_3$ (BFO), has been experimentally realized, leading to a unique route for the stabilization of ferrielectric-like ordering coexisting with an antiferromagnetic state, see \autoref{chemistry} (c) \cite{efe2025nanoscale}. It should be emphasized that, in contrast to the depolarizing field approach, the use of polarizing layers opens avenues for the induction of polar distortions in otherwise non-polar materials. The insertion of magnetic perovskite constituent may, for instance, lead to the creation of unprecedented multiferroic composite materials \cite{fiebig2016evolution,trassin2015low,mundy2016atomically}, in which magnetic and electrical order may coexist and be coupled. Finally, the consideration of other layered frameworks \cite{gradauskaite2025revival}, such as Carpy-Galy \cite{carpy1974systeme,Galy01051974} and Dion-Jacobson \cite{dion_nouvelles_1981,jacobson_interlayer_1985}, may further extend the design of functional superlattices.

\textbf{Ferroelectric switching using crystal chemistry.}
We have so far addressed the use of crystal chemistry and electrostatics at the interface to manipulate ferroelectric ordering in oxide thin films in the as-grown state. Let us now discuss the potential of crystal chemistry control in oxides to act on the polarization state post-growth and achieve polarization reorientation and polarization reversal in the absence of an electric field \cite{efe2024engineering}.
It is shown that polarization-selective chemical bonds may form at the surface of ferroelectric thin films once immersed in a polar solvent. As a result, it is possible to reverse the net polarization of thin films by modifying the surface chemical potential, i.e., most commonly by manipulating the pH \cite{tian2018water}. Crystal chemistry control is also possible within the bulk of ferroelectric layers. Charged defects, such as anion or cation vacancies, affect ferroelectric behavior, and once ordered, they can enforce a preferential polarization direction in thin films \cite{sarott2023controlling,gradauskaite2022ferroelectric,weymann2020full}. In striking contrast to classical interface contributions in polarization reversal, which are structurally defined and remain fixed after design, a charged defect gradient can be manipulated even after the epitaxial design is complete. Taking ferroelectric PTO thin films as a model system, several studies reveal the possibility of poling the films solely using charge defect gradient concentration \cite{wang2009reversible,highland2011equilibrium,stephenson2011equilibrium,mladenovic2025termination}. The oxygen vacancy concentration is controlled via the oxygen partial pressure during the growth and cooldown in various physical vapor deposition processes such as sputtering \cite{weymann2020full}, pulsed laser deposition \cite{sarott2023controlling}, and oxide molecular beam epitaxy \cite{wang2009reversible}. 
Thus, crystal chemistry, either in the form of polarizing charged surface layers, adsorbates, or point-charged defects, which were hitherto considered detrimental, now appears as a new degree of freedom to advance the control of the technology-relevant polarization state \cite{gattinoni2020interface,sarott_multilevel_2022}. They are actively employed to combat current limitations in the loss of polarization in the ultrathin regime and provide an electric-bias-free approach to manipulating net polarization states in thin films. This opens avenues for new applications where physical electrodes are unwanted, such as in the field of catalysis or environmental sensors. The reversibility of surface chemical reconstruction \cite{tian2018water} may also open routes for the design of a new generation of rewritable nanoelectronic architectures.
Next, we will present the recent advances in other means for the manipulation of ferroelectric system lattices and highlight studies revealing the role of mechanical or chemical pressure in the engineering of polarization states.

\section{Acting on polar states using mechanical and chemical pressure}
\label{sec_p}

Mechanical pressure offers a uniquely versatile pathway to manipulate ferroelectric polar textures because it acts on the lattice distortion that underpins ferroelectricity \cite{rabe2007modern,gregg2012stressing}. It may directly alter bond length \cite{andrault1991evolution}, oxygen octahedral tilt patterns \cite{PhysRevLett.105.227203,fowlie2019thickness,lu2015strain}, and spontaneous strain states \cite{lee2000strain,schlom2007strain} in ferroelectric materials and hence trigger structural phase transitions. As a result, mechanical stimuli have emerged as a powerful complementary approach to electrical and chemical methods for tailoring ferroelectric textures across scales. There are various experimental methods for applying pressure to the sample. 
While seminal studies have demonstrated the role of hydrostatic pressure \cite{waqar2015piezoelectric,sani2004high,rouquette2005pressure,ahart2008origin,arnold2010beta,guennou2011multiple} and uniaxial compression \cite{sandvik2023pressure} in the modification of domain patterns or pressure-induced phase transitions in bulk systems, we will focus here on the state of the art related to epitaxial thin films due to their technological relevance and miniaturization potential for electronics devices. Here, the single-crystalline substrate provides mechanical support, and scanning probe microscope (SPM) tips or nanoindenters are employed to deliver localized forces and strain gradients to the sample surface, which modulate the domain structure \cite{lu2012mechanical,wang_mechanically_2020} and trigger phase transitions \cite{edwards2018giant}. When used in combination with the strain and electrostatic engineering empowered by epitaxial thin film growth, the application of pressure at the nanoscale using SPM tips enables the exploration of unconventional phase conversion \cite{muller2025reversible}. Furthermore, the application of pressure is not limited to the mechanical approach but can also be conducted via chemical substitution in the ferroelectric constituent \cite{kan2011composition,wang2021chemical,mumtaz2021chemical,lin2022chemical}. Benefiting from the ionic radii mismatch of the dopant, an internal strain state can be induced and utilized to tune the polar textures in the materials.

    \begin{figure*}[hbt!]
    \includegraphics[width=\columnwidth]{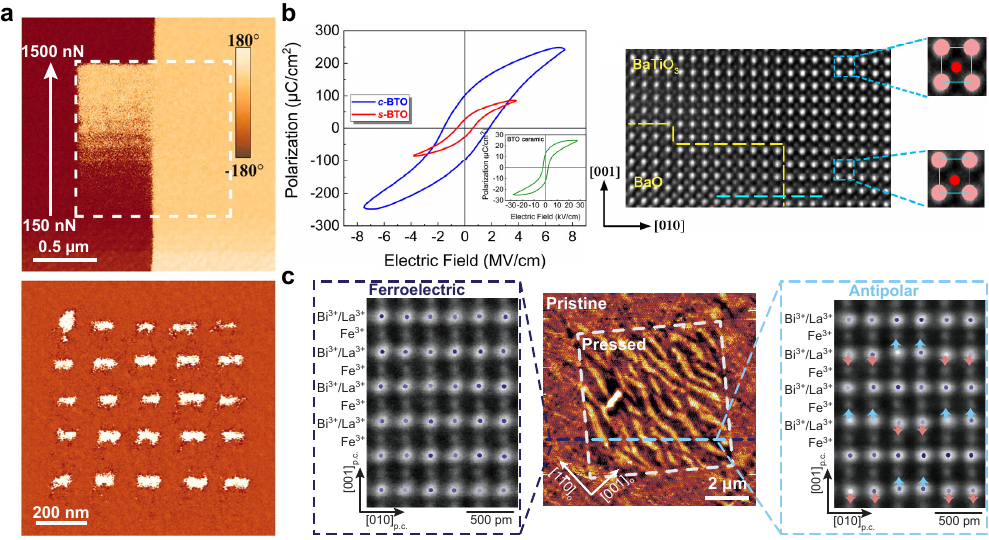}
    \caption{\textbf{Acting on polar states using mechanical and chemical pressure} (a) PFM phase images showing the mechanically induced ferroelectric polarization reversal in BaTiO$_3$ (BTO) thin film. The top panel shows a $1\times1$ \MakeLowercase{\textmu}m$^2$ area (shown by the dashed frame) scanned with the tip under an incrementally increasing loading force on the electrically written bidomain pattern. The bottom panel shows an array of mechanically written nanodomains. Reprinted with permission from Ref.\ \cite{lu2012mechanical}. Copyright 2012, The American Association for the Advancement of Science. (b) Negative chemical pressure in BTO thin film using BTO:BaO composite (c-BTO), the $P$-$E$ loop on the left shows the difference compared with single phase BTO (s-BTO) and BTO ceramic (green), indicating the enhanced ferroelectric polarization by introducing chemical pressure. The cross-sectional STEM image of c-BTO, with two BTO unit cells magnified on the right panel, shows the increase of tetragonality by chemical pressure. Reprinted with permission from Ref.\ \cite{wang2021chemical}. Copyright 2021, American Chemical Society. (c) Ferroelectric-antipolar phase conversion achieved by the synergetic strategy in La-substituted BFO. The AFM image (middle) shows the distinct topography feature of the pristine and pressed region. The HAADF-STEM micrographs of the pristine ferroelectric phase (left) and the stress-induced antipolar phase (right) indicate the pressure-induced phase conversion. Reprinted with permission from Ref.\ \cite{muller2025reversible}. Copyright 2025 by the Authors under Creative Commons Attribution 4.0  International License published by the Nature Publishing Group.
    }
    \label{pressure}
    \end{figure*}

\textbf{Local mechanical pressure application on ferroelectric thin films} Employing SPM tips to apply localized pressure enables the manipulation of ferroelectric polarization and domain structure in thin films. The landmark demonstrations by Lu et al.\ \cite{lu2012mechanical}, revealed how the out-of-plane polarization can be reversed at the nanoscale in BaTiO$_3$ (BTO) thin films solely via the stress gradient induced by an atomic-force microscope (AFM) tip, as shown in \autoref{pressure} (a). 
The existence of a coupling between electrical polarization and strain gradient, referred to as flexoelectricity, supports a polarization reversal induced by local mechanical pressure application \cite{zubko2013flexoelectric}. 
The application of a pressure using an AFM tip in contact with the surface of a thin film leads to a strain gradient in out-of-plane orientation underneath the tip. This induces a local flexoelectric field and a polar bias\cite{lu2012mechanical} that favors only one of the out-of-plane polarization directions.
This seminal work opened avenues for mechanical writing of ferroelectric states even in the nanoscale (\autoref{pressure} (a) bottom panel) for dense data storage.
Follow-up studies have extensively exploited the critical parameters and potential of such a mechanical writing \cite{wang_mechanically_2020} and have expanded its application to various material systems beyond the realm of perovskite ferroelectrics. For example, the mechanical writing of polarization states has been reported in fluorite ferroelectric hafnium-zirconium oxides Hf$_{0.5}$Zr$_{0.5}$O$_{2}$ (HZO) \cite{guan2022mechanical} and ferroelectric polymers \cite{chen2015nonvolatile}.
Most importantly, the potential of flexoelectric-enabled mechanical writing is not limited to collinear polarization configurations, and phase-field simulations have demonstrated that nanoindentation could support the design of stable polar skyrmions and related topological states in PTO thin films \cite{kasai2024mechanical}.

Aside from the flexoelectric-field-driven mechanism, localized pressure application may enable the local modulation of unit cell lattice parameters and, hence, trigger phase transitions. Taking the epitaxial-strain-driven morphotropic phase boundary (MPB) in BFO as a model system \cite{zeches2009strain}, Edwards et al. demonstrated the local control of resistance in the thin films \cite{edwards2018giant}. There, the rhombohedral (R) and tetragonal (T) phases coexist, and while the R constituent can be stabilized upon the application of stress to the thin film via AFM tips, the conversion to a T-phase-dominated structure can be achieved by applying a DC bias to the AFM probe. As a consequence, the populations of these two phases and the conductive boundaries between them can be reversibly tuned through a combined stress and electric field method.

\textbf{Chemical pressure in ferroelectric thin films} 
In addition to conventional mechanical pressure, chemical pressure has emerged as a powerful lever for engineering ferroelectricity \cite{lin2022chemical}. By substituting the constituent elements with species exhibiting different ionic radii, an internal strain state can be built up in the oxide thin films. Moreover, the applied chemical pressure can be tuned both by changing the dopant species and the substitution level. Taking the systematic study of rare-earth-doped BFO thin films as an example, it is demonstrated how decreasing the $A$-site ionic radius drives structural evolution across MPBs \cite{kan2011composition}. 
Strikingly, the utilization of chemical pressure allows for the achievement of negative pressure conditions by introducing chemical species of large size. The pioneering work of Wang et al.\ \cite{wang2021chemical} demonstrated that introducing BaO into a BaTiO$_3$ thin film matrix led to coherently imposed negative chemical pressure and extraordinarily high tetragonality, as illustrated in \autoref{pressure} (b). The left figure shows the comparison of hysteresis loops among the BTO:BaO composite (c-BTO), single-phase BTO (s-BTO), and bulk BTO ceramic (inset green curve). It is shown that c-BTO exhibits a large spontaneous polarization that vastly exceeds the bulk and single-phase values. Chemical pressure can also be introduced by light element implantation, such as helium (He), and enhances the ferroelectricity in both conventional perovskite oxide ferroelectrics \cite{herklotz2025polarization} and HfO$_2$-based ferroelectrics \cite{kang2022highly}.
These advances highlight chemical pressure as an electric-field-free means of controlling polar states.

\textbf{Synergetic combination of chemical substitution and mechanical pressure for polar to antipolar phase conversion} 
The combination of microscopic pressure and chemical engineering can further facilitate the modulation, or even the creation, of functionalities in oxide thin films. Taking advantage of the rich phase diagram of multiferroic BFO, cationic substitution can bring the system to the verge of a polar to antipolar phase transition \cite{mundy2022liberating,caretta2023non,dedon2018strain}. There, a local mechanical stimulus may suffice to trigger the phase conversion. This has been demonstrated in the La-substituted BFO system \cite{muller2025reversible} (\autoref{pressure} (c)). We note, in addition, that in the BFO-based system, the $A$-site Bi cation is responsible for long-range ferroelectric ordering via the lone pair mechanism \cite{muller2023ferroelectric,fiebig2016evolution,muller2021training}. For a well-defined La concentration, the application of localized pressure on the films stabilizes a ferroelectric to an antipolar phase transition and leads to the complete elimination of polarization in the mechanically stressed area. Significantly, the ferroelectric polarization in the pressed region can be fully recovered by applying a continuous bias, even at the nanoscale \cite{muller2025reversible}. Such reversible control over the polar-antipolar phase transition thus empowers access to multiple levels of polarization states for future beyond-binary response electronics. This proof of principle and synergetic modulation of ferroelectricity motivates further exploration into the combined effects of mechanical force, strain gradient, electrostatic boundary conditions, and cationic substitution on the final polar textures. Hence, chemical engineering of ferroelectric lattices, jointly with mechanical pressure application, enables going beyond the realm of manipulation of polarization states allowed with solely electric field application. 
So far, we have discussed moderately invasive routes for the control of ferroelectric polarization in thin films at zero volts. Let us now consider a remote and non-invasive approach and focus on light to act on ferroelectric states.


\section{Optical control of polarization states in ferroelectric thin films}
\label{sec_l}

Light-matter interaction facilitates a novel electrode-free handle on modifying polarization states in ferroelectric materials via a variety of elasto- and electro-optical effects. In analogy to the magnetic counterpart, where the magnetization can be reversibly switched \cite{lambert2014all,manz2016reversible} and modulated in the ultrafast timescale using pulsed laser \cite{duong2004ultrafast}, recent advances have demonstrated that ferroelectric polarization can also be remotely controlled via light illumination. 
Such optical control can originate from various mechanisms based on different photo-induced effects, such as flexoelectric, bulk photovoltaic, and photostrictive effects. Here, we will focus on the modification of ferroelectric polarization in epitaxial thin films by light-induced strain gradients, charge carrier generation, and accumulation. Note that due to the highly strained and clamped nature of the epitaxial thin film systems considered here, we will not cover photostriction in this perspective. Nevertheless, we direct the reader to other sources \cite{chen2021photostrictive, matzen2019tuning,kundys2010light}.

    \begin{figure*}[hbt!]
    \includegraphics[width=\columnwidth]{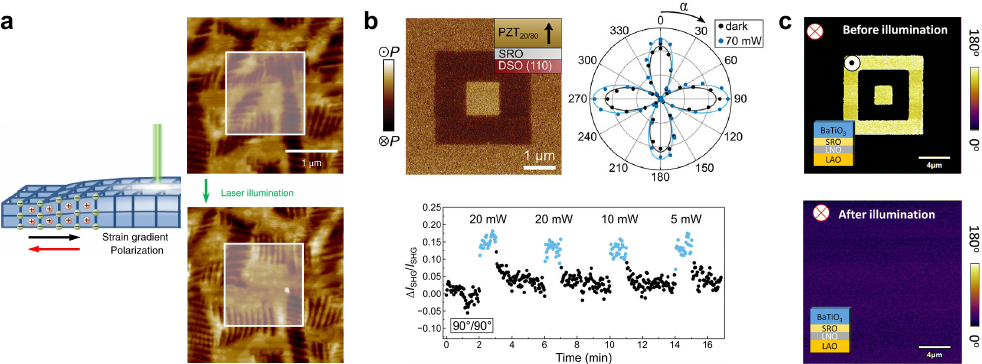}
    \caption{\textbf{Optical control of polarization states in ferroelectric thin films} (a) Light induced flexoelectricity. Topography image of an as-grown mixed-phase BFO thin film. The white square indicates the illuminated area, the same area after light illumination shows a clear phase redistribution of T-BFO and R-BFO phases. Reprinted with permission from \cite{liou_deterministic_2019}. Copyright 2019, The Author(s), under exclusive licence to Springer Nature Limited. (b) Transient light-induced polarization enhancement. VPFM image of the single-domain PZT$_{20/80}$ thin film after box-in-box poling. Second harmonic generation (SHG) polarizer measurement on the upward-polarized film for fixed $p$-polarized SHG light before (black) and during UV-light exposure (blue). The solid lines represent the fits to the $4mm$ point group. SHG time trace under repeated UV-light exposure for $p$-polarized incident probe light and detected SHG light. Reprinted with permission from Ref.\ \cite{Sarott_reversible}. Copyright 2024 by the Authors under Creative Commons Attribution 4.0  International License published by Wiley-VCH. (c) Domain erasure by combining the optical response and imprint. PFM phase images collected before and after illumination for the BTO/SRO sample. Reproduced from Ref.\cite{tan2022control} with permission from the Royal Society of Chemistry.
    }
    \label{light}
    \end{figure*}

\textbf{Light-induced flexoelectric effect} 
An increasing number of reports highlight the use of light as an effective stimulus to modulate ferroelectricity or even multiferroicity in thin films \cite{Recent_progress}. The interaction between light and polar materials introduces a range of intriguing physical phenomena that can potentially be employed for remote and dynamic control over polarization. The seminal work by Li et al.\ demonstrated a giant optical enhancement of the aforementioned flexoelectric effect in BFO thin films \cite{li2015giant}. Here, the light illumination induces substantial lattice expansion via transient thermal \cite{liou_deterministic_2019} or optical rectification \cite{liou2021extremely} effects. This leads to a redistribution of the strain gradient and subsequent modification of the flexoelectric polarization. Taking advantage of both the light-induced strain gradient and the notorious sensitivity at the MPB, as discussed in section \ref{sec_p}, light illumination gives rise to deterministic control over the domain patterns in BFO thin films at the strain-driven MPB \cite{liou_deterministic_2019} (\autoref{light} (a)).
Most importantly, the flexoelectric effect connects light-induced mechanical strain and electrical polarization.
These works establish light as a powerful, non-invasive tool to manipulate ferroelectric polarization in thin films.

\textbf{Bulk photovoltaic effect}
The ferroelectric polarization states can also be modulated by light illumination via the bulk photovoltaic (BPV) effect due to the non-centrosymmetric nature of ferroelectric materials. When an above-bandgap light source is employed, the lack of inversion symmetry leads to an asymmetric distribution of photo-excited charge carriers, resulting in a direct electric current in the material under illumination, which then gives rise to local space-charge fields \cite{Fridkin_BPE}. These fields then facilitate the modification of ferroelectric polarization, such as photoinduced polarization reorientation. For instance, these space-charge fields are sufficient to reversibly switch ferroelectric domains in BFO, without the necessity of applied external fields \cite{Yang_BVE}. Another noteworthy illustration of the ferroelectric polarization reorientation, facilitated by the BPV, has been reported in a study using the tip-enhanced photovoltaic effect \cite{PhysRevB96134107, Yang_BV}. At the location of the tip contact, the photocurrent density is increased, thereby producing a local internal electric field that exceeds the coercive field of the material, enabling local switching \cite{Yang_BV}. 
Furthermore, the manipulation of the polarization of incoming light paves new avenues for further control over ferroelectric polarization switching. 
It is demonstrated that, in BFO thin films, the circular BVP effect strongly depends on the helicity of the incident light \cite{knoche2021anomalous}. 
Such an observation points to the potential for more precise control over the optical manipulation of polarization states.

\textbf{Optically induced modulation of interfacial electrostatic boundary conditions for the manipulation of ferroelectric polarization}
In technology-relevant ultrathin films, the electrostatic boundary conditions set by the interfaces play a pivotal role in manipulating ferroelectric polarization. The utilization of above-bandgap UV light illumination enables the generation of photo-excited charge carriers within the system. These charge carriers undergo separation due to the space charge electric field, thereby inducing charge accumulation towards the electrode interface, where the Schottky barrier is formed. This results in the modification of the bound charge screening condition, and a transient enhancement (or suppression) of the spontaneous polarization under UV illumination can thus be triggered in PTO thin films (\autoref{light} (b)) \cite{Sarott_reversible}. The nature of the Schottky barrier between the electrode and the ferroelectric material depends on the carrier type in the polar layer, and the band bending is drastically influenced by the polarization direction \cite{pintilie2005metal}. The ferroelectric bound charge accumulation may, for instance, enhance or attenuate the internal field and corresponding band bending at the Schottky interface, depending on the charge carrier type in the ferroelectric material and on the polarity of the bound charge at the interface with the metallic buffer. Therefore,  the transient optical modulation of the polarization in ferroelectric thin films is materials-dependent and ferroelectric polarization orientation-dependent \cite{Sarott_reversible}. Importantly, starting from a multidomain state, this optical process can lead to remanent optical poling and the stabilization of a single domain state after light exposure. Acting on the electrostatic boundary conditions of ferroelectric oxide thin films through combined UV illumination and depolarizing field tuning appears to be a critical step towards advanced optical control of spontaneous polarization states. Furthermore, the optical erasure of written domains can be achieved by leveraging the influence of photo-excited charge carriers on the electric imprint field ($E_{\text{imp}}$). Such an internal field, which sets a preferred polarization direction, arises from asymmetries in the film system, such as unequal work functions of top and bottom electrodes, defects or charge trapping at interfaces, or chemical asymmetry \cite{Long_Nature_intro, Long_Nature_intro2, Long_Nature_intro3, Long_Nature_intro4, Long_Nature_intro5}. 
Tan et al.\ revealed that tailoring the bottom electrodes underneath BTO thin films can lead to selective up-to-down or down-to-up polarization switching under light exposure in the absence of top electrodes \cite{tan2022control} (\autoref{light} (c)). Here, the photo-induced charge carriers transiently screen the polarization, and as the light source is removed, the polarization rearranges along the favored direction set by the imprint field. Such a phenomenon enables the erasure of artificially written domains (i.e., characterized by a polarization pointing against the direction set initially by the imprint field) and the control over it by selecting the appropriate bottom electrode material.
This optical control is not limited to single-layer films but also provides an electric-bias-free means of controlling ferroelectric-based devices with top electrodes, such as ferroelectric tunnel junctions. There, light-induced changes in polarization occur due to photoexcited charge carriers and lead to a change in the junction's resistance \cite{Long_Nature}.
Together, the aforementioned advances shed light on non-invasive, potentially ultrafast control over ferroelectric-based oxide electronics. Thus, in the context of the current rush for ultrafast electronics \cite{li2004ultrafast,ma2020sub}, neuromorphic computing platforms \cite{chanthbouala2012ferroelectric,oh2019ferroelectric,mikolajick2023ferroelectric}, and dynamically programmable electronic interfaces \cite{zhang2024reconfigurable}, investigations on optical manipulation of ferroelectric polarization have become particularly timely.
In the next section, we will discuss the accompanying challenges and opportunities relevant to the unconventional modulation of ferroelectric states.

\section{Upcoming challenges and opportunities for unconventional modulation of polarization in thin films}

The aforementioned new approaches for tuning ferroelectric polarization have opened up new possibilities for designing ferroelectric-based devices. However, questions regarding the robustness and speed of switching events remain underexplored. We will discuss the recent attempts to overcome the notorious ferroelectric fatigue and advances towards faster memory devices using the methods we discussed above.

Ferroelectric polarization fatigue is a phenomenon that denotes the deterioration of the switching ability of a ferroelectric material, typically in a capacitor structure, after it has been subjected to multiple polarization reversals \cite{tagantsev2010domains}. It has been one of the main limiting factors for long-term, robust ferroelectric-based devices. The redistribution and accumulation of mobile charged defects, such as oxygen vacancies, can pin the polarization direction in certain regions as well as the domain wall motion with the increasing cycling number, resulting in being the predominant origins of ferroelectric fatigue \cite{lou2009polarization,baek2011nature}. The crystal chemistry approach may be the key to preventing the degradation of polarization switchability. The Aurivillius layer structure, for instance, converts off-stoichiometry charged defects into structural defects and hence exhibits fatigue-free behavior in both bulk \cite{de1995fatigue} and thin films \cite{gradauskaite2020robust}.
However, the dominant in-plane polarization component in Aurivillius compounds makes their technological integration challenging since the out-of-plane geometry for capacitors is generally preferred. Therefore, the incorporation of perovskites might enable the combination of the inherently low fatigue of the layered oxides with a remanent out-of-plane polarization, making these materials a suitable platform for robust devices. 
As for the pressure approach, it is reported that the resistance to electric-poling-induced fatigue can be significantly enhanced via the application of strain \cite{lee2012ferroelectric}. However, the role of fatigue under mechanical writing remains underexplored and calls for studies on lattice degradation and possible charge defect accumulation at created structural defects, such as out-of-phase boundaries and dislocations. 
Finally, the effect of light illumination on electric-poling-induced fatigue resistance and the establishment of all-optical ferroelectric switching call for further investigation. Preliminary results on the optical modulation of ferroelectric polarization point out that the photo-driven redistribution of charge carriers also faces the diffusion limit, which can lead to the pinning of polarization and tunable fatigue behavior \cite{Sarott_reversible}.

From the perspective of switching speed, the optical control of polarization undoubtedly holds the most significant potential for ultrafast control of polar states. Recent research advances have revealed the potential of manipulating ferroelectric domain structures or even tuning the polar phases on the ps or even fs time scale by employing pulsed lasers as the external stimuli \cite{liou2021extremely,guzelturk2023sub}.
Besides visible and UV light, other frequency domains, particularly THz \cite{chen2016ultrafast,basini_terahertz_2024}, have revealed the possibility of robust control over ferroelectric polarization or even the induction of dynamic polarization. In addition, the recent boost in the development of freestanding membranes \cite{ji2019freestanding,lu2016synthesis,pesquera2020beyond,dong2019super,cai2022enhanced, chiabrera2022freestanding, pryds2024twisted,brand2025strain}, which are released from rigid substrates and liberated from the clamping effects, has paved a new pathway for the synergistic combination of epitaxial strain release and optical control. Seminal studies have highlighted the timescale at which the light-induced charge carriers can influence metastable electric dipole ordering in thin film heterostructures. Subsecond light exposure has been reported to enable domain erasure in freestanding ferroelectric membranes \cite{pal2025subsecond}. Nevertheless, more information and time-resolved characterizations of these phenomena are needed to assess the competitiveness of such an approach compared to magnetically ordered systems in which the magnetic state can be accessed on the fs timescales.

\section{Summary and Outlook}

In summary, we highlight the recent breakthroughs in tailoring ferroelectric polarization in epitaxial thin films through electrical-bias-free methods, emphasizing approaches that utilize crystal chemistry, local pressure, and light illumination. Firstly, we provide insights regarding the developments in employing both depolarizing and polarizing fields for designing electrical dipole textures. By engineering the crystal chemistry, not only can the polarization direction be deterministically controlled, but complex polar structures and even unprecedented functionalities can also be designed artificially. Moreover, electrostatic engineering may provide additional degrees of freedom for electrode-free ferroelectric switching. We then address the impact of local mechanical pressure application on ferroelectric poling and the manipulation of polar phase architectures through both physical contact and chemical substitution. A synergetic approach that combines mechanical pressure and chemical engineering facilitates continuous control of the net polarization state in thin films through pressure-driven polar-antipolar phase conversions, which opens new pathways beyond binary responses. Finally, we elaborate on the optical control of polarization in ferroelectric thin films through various mechanisms. The ferroelectric polarization can be remotely modulated and even switched under light illumination by acting on the strain state or the electrostatic boundary conditions due to charge carrier generation. It is noteworthy that, in combination with the recent boom in freestanding membranes, the above strategies will further broaden the possibilities of manipulating ferroelectricity with electrical-bias-free means and promote innovation in ferroelectric-based applications.

\begin{acknowledgments}
All authors acknowledge the Swiss National Science Foundation (SNSF) under project no.\ 200021--236413 and 200021--231428. M.T.\ and B.Y.\ acknowledge the ETH Zurich Research Grant funding under reference 22--2 ETH-016. M.T.\ acknowledges the Swiss National Science Foundation Spark funding CRSK--2--227378. All authors thank Oskar Leibnitz and Christoph Dr{\"a}yer for proofreading the manuscript.
\end{acknowledgments}

\section*{Conflict of Interest}

The authors declare no conflict of interest.

\section*{Keywords}
Ferroelectric, thin films, crystal chemistry, flexoelectricity, optical control

\newpage
\bibliography{REFs}
\end{document}